\documentclass[aps, prb, twocolumn, amssymb, amsmath, showpacs, superscriptaddress]{revtex4-1}
\usepackage{bm}
\usepackage{times}
\usepackage{graphicx}
\usepackage{color}
\usepackage{dcolumn}
\usepackage[colorlinks=true, letterpaper=true, pdfstartview=FitV, linkcolor=blue, citecolor=blue, urlcolor=cyan]{hyperref}
\usepackage{appendix}
\usepackage[normalem]{ulem}

\newcommand{\vect}[1]{\boldsymbol{#1}}                 
\usepackage{helvet}
\DeclareMathOperator{\sech}{sech}

\begin{document}

\title{Light-induced thermal noise \textit{anomaly} governed by quantum metric}
\author{Longjun Xiang}
\affiliation{College of Physics and Optoelectronic Engineering, Shenzhen University, Shenzhen 518060, China}
\author{Lei Zhang}
\email[]{zhanglei@sxu.edu.cn}
\affiliation{State Key Laboratory of Quantum Optics and Quantum Optics Devices, Institute of Laser Spectroscopy,
Shanxi University, Taiyuan 030006, China}
\author{Jun Chen}
\affiliation{State Key Laboratory of Quantum Optics and Quantum Optics Devices, Institute of Theoretical Physics, Shanxi University, Taiyuan 030006, China}
\author{Fuming Xu}
\affiliation{College of Physics and Optoelectronic Engineering, Shenzhen University, Shenzhen 518060, China}
\author{Yadong Wei}
\affiliation{College of Physics and Optoelectronic Engineering, Shenzhen University, Shenzhen 518060, China}
\author{Jian Wang}
\email[]{jianwang@hku.hk}
\affiliation{College of Physics and Optoelectronic Engineering, Shenzhen University, Shenzhen 518060, China}
\affiliation{Department of Physics, University of Hong Kong, Pokfulam Road, Hong Kong, P. R. China}

\date{\today}

\begin{abstract}
Traditionally, thermal noise in electric currents, arising from thermal agitation,
is expected to increase with temperature $T$ and disappear as $T$ approaches zero.
Contrary to this expectation, we discover that the resonant DC thermal noise (DTN)
in photocurrents not only persists at $T=0$ but also exhibits a divergence proportional to $1/T$.
This thermal noise \textit{anomaly} arises from the unique electron-photon interactions
near the Fermi surface, manifesting as the interplay between the
inherent Fermi-surface property and the resonant optical selection rules of DTN,
and thereby represents an unexplored noise regime.
Notably, we reveal that this \textit{anomalous} DTN, especially in time-reversal-invariant systems,
is intrinsically linked to the quantum metric. We illustrate this \textit{anomalous} DTN in massless Dirac materials,
including two-dimensional graphene, the surfaces of three-dimensional topological insulators,
and three-dimensional Weyl semimetals, where the quantum metric plays a pivotal role. 
Finally, we find that the total noise spectrum at low temperatures,
which includes both the DC shot noise and the \textit{anomalous} DTN,
will universally peak at $\omega_p=2|\mu|$ with $\omega_p$ the frequency of light and
$\mu$ the chemical potential of the bulk crystals.
\end{abstract}

\maketitle

\noindent{\textit{\textcolor{blue}{Introduction.}}} ---
Quantum fluctuations, manifesting as quantum noise,
are ubiquitous in transport processes and are well understood in mesoscopic conductors \cite{Datta1, Datta2, Buttiker}.
Among these, shot noise (SN) in mesoscopic systems,
arising from charge quantization, is considered the dominant source of noise at low temperatures
\cite{Buttiker, shot2003}.
Oppositely, thermal noise (TN), driven by thermal agitation,
becomes the prevailing source of quantum noise at high temperatures
and disappears at low temperatures \cite{Buttiker, Nyquist}.

Beyond mesoscopic conductors, the SN of electric current in bulk crystals under a static electric field
is closely tied to the quantum metric of Bloch electrons
\cite{quantumgeometry}.
Meanwhile, the TN of current in similar scenarios can manifest as an intrinsic Hall signature,
even in time-reversal-invariant systems \cite{WeiPRL}.
Additionally, the DC shot noise (DSN) of photocurrent in bulk crystals under optical electric fields
has been developed as a tool to probe the quantum geometric properties of centrosymmetric quantum materials \cite{NagaosaSN1, xiangDSN}.
However, the DC thermal noise (DTN) of photocurrent in bulk crystals under optical electric fields remains unexplored,
leaving a critical gap in understanding the quantum noise of bulk crystals under light irradiation.

In this \textit{Letter}, we develop the quantum theory of DTN
for photocurrent in bulk crystals under an optical electric field.
We demonstrate that the DTN in time-reversal-invariant systems is intimately related to the quantum metric,
which as the quantum geometric origin of the intrinsic nonlinear Hall effect
recently received significant attention
\cite{intrinsic20211, intrinsic20212, XuSY2023, Wang2023, MNHE1, MNHE2, QMNP, Jia2024}.
Remarkably, unlike the TN occurred in mesoscopic conductors and in bulk crystals under static electric fields,
we discover that the DTN not only persists at zero temperature ($T=0$) but also can exhibit a divergent $1/T$ behavior.
This thermal noise \textit{anomaly} arises from
the distinctive electron-photon interactions near the Fermi surface,
essentially due to the interplay between the inherent Fermi-surface property
and the optical selection rule of DTN.
This phenomenon thus unveils a novel quantum noise regime.
To illustrate our proposals, we investigate the \textit{anomalous} DTN in quantum materials
with a nontrivial quantum metric, including two-dimensional (2D) graphene
\cite{graphene2004, grapheneYBZhang, graphenerise, Xiao2007graphene},
the surface of three-dimensional (3D) topological insulators \cite{Bi2Te3, ModelHam, Fu2009, Kane2010, XLQi2011, JEmooreTI},
and 3D Weyl semimetals \cite{Weyl1, Weyl2, Weyl3}.
Finally, we show that the total noise spectrum at low temperatures,
which contains both the DSN and the \textit{anomalous} DTN, will universally peak at $\omega_p=2|\mu|$,
where $\omega_p$ is the frequency of optical electric fields
and $\mu$ is the chemical potential of the bulk crystals.

\bigskip
\noindent{\textit{\textcolor{blue}{Unique electron-photon interaction.}}}
---
In mesoscopic physics, it is well established that the TN of electric current vanishes
at zero temperature. Specifically, in a two-terminal mesoscopic conductor,
the TN, applicable in both linear and nonlinear regimes,
can be expressed as \cite{Buttiker, fnn} ($e=\hbar=k_B=1$)
\begin{align}
S_T
=
\frac{1}{\pi} \sum_{\alpha,n} \int_E f_{\alpha}(1-f_\alpha) T_n
=
-\frac{T}{\pi} \sum_{\alpha,n} \int_E f_{\alpha}' T_n,
\label{STmeso}
\end{align}
where $\int_E=\int dE$,
$\alpha=\{L,R\}$ denotes the left and right electrodes,
$T_n=T_n(E)$ stands for the transmission of the $n$th channel,
$f_\alpha \equiv f(E,\mu_\alpha)$ represents the equilibrium Fermi distribution function
(Here $E$ is the energy and $\mu_\alpha$ is the chemical potential of the electrode $\alpha$),
and $f_\alpha' \equiv \partial f_\alpha/\partial E$.
The presence of $f_\alpha'$ indicates that
the TN is contributed by the electrons on the Fermi surface \cite{Haldane}.
As a result, at $T=0$ we find $f_\alpha' = -\delta \left( \mu_\alpha - E \right)$
and we arrive at
\begin{align}
\lim_{T \rightarrow 0}S_{T} = \lim_{T \rightarrow 0} \frac{T}{\pi}  \sum_{\alpha,n} T_n(\mu_\alpha) = 0
\label{zero}
\end{align}
since $\sum_{\alpha,n} T_n(\mu_\alpha)$ is finite.

In bulk crystals, the noise of photocurrent generally is given by \cite{xiangDSN}
\begin{align}
\bar{S}_{T}=
\dfrac{1}{2}\sum_{mn} \int_k J^a_{mn}J^b_{nm}
\left(
f_{nm}^2
+
\bar{f}_{nm}
\right).
\label{Sab}
\end{align}
Here $\int_k=1/V\int d\vect{k}/(2\pi)^d$ with $d$ and $V$ being the spatial dimension and the volume of the system \cite{Sipe2000},
respectively;
$J^a_{nm}$ is the matrix element of photocurrent operator;
$f_{nm}=f_n-f_m$ and $\bar{f}_{nm}=f_n(1-f_n)+f_m(1-f_m)$,
where $f_n \equiv f(\epsilon_n, \mu)$ represents the equilibrium Fermi distribution function
but with $\epsilon_n$ being the energy of the $n$th Bloch band and $\mu$ being the chemical potential of the bulk crystal.
In Eq.~(\ref{Sab}), the first term featuring the Fermi-sea property gives the DSN \cite{xiangDSN}
while the second term gives the DTN,
which behaves as the Fermi-surface property like Eq.~(\ref{STmeso})
due to $f_n(1-f_n)=-T\partial f_n/\partial \epsilon_n=-Tf_n'$.

At the second order of optical electric fields and within the two-band limit \cite{NagaosaSN1,twobandlimit1},
the resonant DTN of photocurrent in bulk crystals has two distinct contributions
and can be formally expressed as [see Eqs.~(\ref{etaL1}-\ref{etaL2}) derived below]:
\begin{align}
\bar{S}_{T}^{(1)} &=
-T \int_k
\left(f_{n}'+f_{m}'\right)
N_{nm}^{(1)} \delta(\omega - \omega_{mn}),
\label{DTN1}
\\
\bar{S}_{T}^{(2)} &=
-
T
\int_k
\omega
\left(
f_{n}''
-
f_{m}''
\right)
N_{nm}^{(2)}
\delta(\omega - \omega_{mn}),
\label{DTN2}
\end{align}
where $f_{n}'' \equiv \partial^2 f_{n}/\partial \epsilon_n^2$,
$\omega$ is the frequency of the optical electric field,
$\hbar\omega_{mn}=\epsilon_{mn}=\epsilon_m-\epsilon_n$,
and $N_{nm}^{(1/2)}$ related to \textit{quantum metric} will be calculated below.
Compared to the TN described by Eq.~(\ref{STmeso}),
besides the delta function given by the inherent Fermi-surface property
$f_{n}'$ and $f_{n}''$ at zero temperature,
we notice that the DTNs $\bar{S}_T^{(1/2)}$ contain an additional delta function
$\delta(\omega-\omega_{mn})$ particularly due to the optical selection rule \cite{transitionrule}.
Consequently, the conclusion implied by Eq.~(\ref{zero}) is no longer hold
since $\lim_{T\rightarrow 0} T \int_k f_n' =0$ 
but $\lim_{T\rightarrow 0} T f_n' \neq 0$
and $\lim_{T\rightarrow 0} T f_n'' \neq 0$
when the $k$-integral is killed by the additional optical selection rule.

\begin{figure}[t!]
\includegraphics[width=8.0cm]{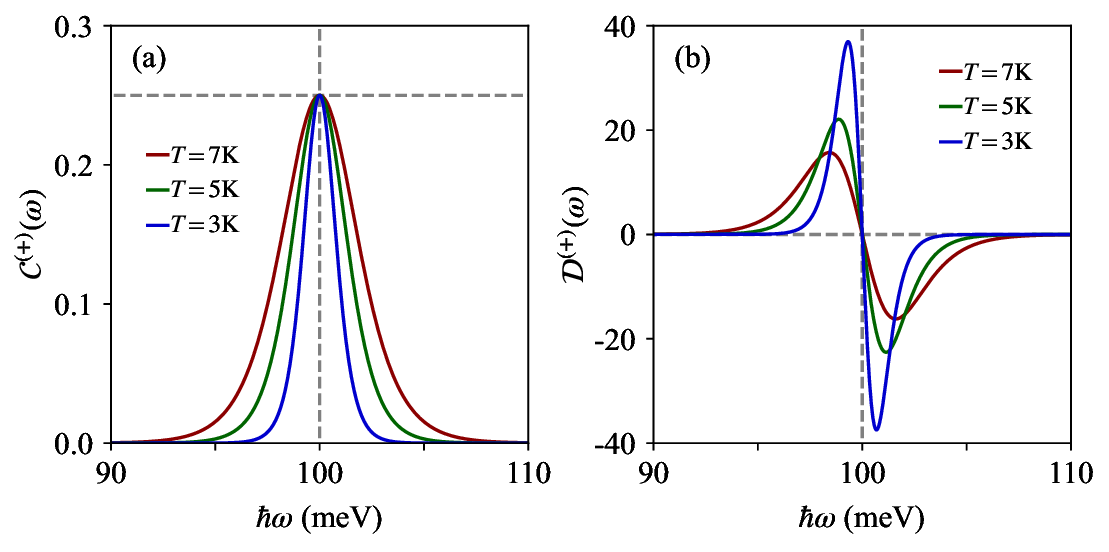}
\caption{
The temperature and frequency dependencies of
(a) $\mathcal{C}^{(+)}$ and (b) $\mathcal{D}^{(+)}$,
as calculated in Eqs.~(\ref{first}-\ref{second}), respectively.
Here we have set $\mu=50 \mathrm{meV}$ so that
the resonant peaks appear near $\omega=2\mu=100 \mathrm{meV}$.
}
\label{FIG1}
\end{figure}

Specifically, we take the two-band Dirac Hamiltonian into account \cite{DiracEQ, Vafek2014}
\begin{align}
H=\vect{d}\cdot\vect{\sigma} \label{Hamiltonian},
\end{align}
whose band dispersions are given by $\epsilon_{\pm}=\pm d$,
where $+(-)$ correspond to the upper (lower) band, respectively,
and $d^2=|\vect{d}|^2=\sum_{i} d_i^2$ with $d_i$ being the functions of $\vect{k}$.
For this Hamiltonian, Eqs.(\ref{DTN1}-\ref{DTN2}) can be simplified to
\begin{align}
\bar{S}_{T}^{(1)}
=
\mathcal{C} (\omega)
\mathcal{N}^{(1)},
\qquad
\bar{S}_{T}^{(2)}
=
\mathcal{D}(\omega)
\mathcal{N}^{(2)},
\label{simp}
\end{align}
where $\mathcal{N}^{(1/2)} \equiv \int_k N_{-+}^{(1/2)}\delta(\omega-2d)$
are finite and independent of temperature,
$\mathcal{C}(\omega)=\mathcal{C}^{(-)}(\omega)+\mathcal{C}^{(+)}(\omega)$
with $\mathcal{C}^{(\pm)}(\omega) \equiv - T f_{\pm}'$ \cite{energy} and
$\mathcal{D}(\omega)=\mathcal{D}^{(-)}(\omega)-\mathcal{D}^{(+)}(\omega)$
with $\mathcal{D}^{(\pm)}(\omega) \equiv -\omega T f_{\pm}''$.
Explicitly, they are given by \cite{sup}
\begin{align}
\mathcal{C}^{(\pm)}
&=
\frac{1}{4}
\sech^2
\left( \dfrac{\hbar \omega \mp 2 \mu }{4k_BT} \right),
\label{first}
\\
\mathcal{D}^{(\pm)}
&=
\frac{\mp\hbar\omega}{4k_BT}
\sech^2\left(\frac{\hbar \omega \mp 2 \mu }{4k_BT}\right)
\tanh\left(\frac{\hbar \omega \mp 2 \mu }{4k_BT}\right),
\label{second}
\end{align}
which are dimensionless universal functions
of temperature, frequency, and chemical potential.
Here $\hbar$ and $k_B$ are restored by dimension analysis.
Note that $\mathcal{C}^{(\pm)}(\omega)$ exhibits a resonant peak at $\omega=2|\mu|$,
with a universal maximum value $1/4$,
even in the limit as $T \rightarrow 0$,
as shown in Fig.~\ref{FIG1}a for $\mathcal{C}^{(+)}(\omega)$.
Consequently, $\bar{S}_T^{(1)}$ defines the first type of \textit{anomalous} DTN.
Furthermore, $\mathcal{D}^{(\pm)}(\omega)$ obtained by taking the derivative of $\mathcal{C}^{(\pm)}(\omega)$,
gives two resonant peaks
on either side of $\omega=2|\mu|$, even in the limit as $T \rightarrow 0$,
as shown in Fig.~\ref{FIG1}b for $\mathcal{D}^{(+)}(\omega)$.
As a result, $\bar{S}_T^{(2)}$ defines the second type of \textit{anomalous} DTN,
characterized by a temperature dependence of $1/T$ [see Eq.~(\ref{second})].

To close this section, we wish to conclude that the \textit{anomalous} DTNs 
arise from the unique electron-photon interactions near the Fermi surface,
which manifests as the interplay between the inherent Fermi-surface property $f_n'$ or $f_n''$
and the resonant optical selection rule $\delta(\omega-\omega_{mn})$ of DTN
---
a phenomenon that can not be expected in mesoscopic conductors or in bulk crystals under a static electric field.
Therefore, this thermal noise \textit{anomaly} represents a novel quantum noise regime beyond the conventional thermal noise.
Building on these qualitative insights, we proceed to develop the quantum theory for resonant DTN.

\bigskip
\noindent{\textit{\textcolor{blue}{Quantum theory of DTN.}}} ---
Similar to the DSN, the lowest-order DTN occurring at the second order of
the optical electric field $\vect{E}$ is given by \cite{xiangDSN}:
\begin{align}
\bar{S}_T^{(2)}
=
\delta(\Omega_1)
\left(\bar{\sigma}_{L/C} + \tau \bar{\eta}_{L/C} \right) \delta_{L/C},
\label{DTNdecomp}
\end{align}
where $\Omega_1$ represents the response frequency for $t-t'$,
$\tau$ is the effective illumination time.
The tensors $\bar{\sigma}$ and $\bar{\eta}$ denote the response coefficient for the shift and injection DTN, respectively.
Here the shift (injection) DTN characterizes the quantum fluctuation of the shift (injection) photocurrent operator.
Additionally, $\delta_L \equiv |\vect{E}|^2$ and $\delta_C \equiv |\vect{E}\times\vect{E}^*|$
correspond to linearly polarized light (LPL) and circularly polarized light (CPL), respectively.
As indicated by the subscripts $L$ and $C$ of
$\bar{\sigma}$/$\bar{\eta}$ in Eq.(\ref{DTNdecomp}),
we notice that both $\bar{\sigma}$ and $\bar{\eta}$ can be excited either by LPL or CPL,
similar to the DSN \cite{xiangDSN} and the DC photocurrent \cite{HWang2020, JEMoore}.

In time-reversal-invariant systems, which are the focus of this \textit{Letter},
only the shift/injection DTN excited by CPL/LPL can appear \cite{Todd}.
The shift DTNs excited by CPL feature the antisymmetric form,
which disappears for the linear Dirac Hamiltonian \cite{sup}.
Consequently, we will concentrate on the injection DTN in the following.
Specifically, the response tensors for injection DTN
within the two-band limit are given by \cite{sup}
\begin{align}
\bar{\eta}^{abc}_{L,1}
&=
-
T
\int_k \left( f_{n}' + f_{m}' \right)
\Xi^{abc}_{nm}
\delta(\omega-\omega_{mn})
,
\label{etaL1}
\\
\bar{\eta}^{abc}_{L,2}
&=
-T
\int_k
\omega
\left( f_{n}'' -f_{m}'' \right) \Lambda^{abc}_{nm}
\delta(\omega-\omega_{mn}),
\label{etaL2}
\end{align}
where $\Xi^{abc}_{nm} = - \pi \left( \Delta^a_{nm} \right)^2 g^{bc}_{nm}/2$
with $g^{bc}_{nm}=r^b_{nm}r^c_{mn}+r^c_{nm}r^b_{mn}$ being the quantum metric
(Here $r^b_{nm}$ is the interband Berry connection) and
$\Delta^a_{nm}=v_{n}^a-v_{m}^a$ 
(Here $v_{n}^a \equiv \partial \epsilon_n /\partial k_a$ is the group velocity for the $n$th Bloch band).
In addition,
$\Lambda^{abc}_{nm}= \pi \Delta^a_{nm}\left( g^{ac}_{nm} v_{n}^b  + g^{ab}_{nm} v_{n}^c \right)/4$.

Several observations are in order.
First, we note that $\bar{\eta}^{abc}_{L,1}$ and $\bar{\eta}_{L,2}^{abc}$
give rise to the first and second type of \textit{anomalous} DTNs, respectively,
resulting from the interplay between the Fermi-surface property and the optical selection rule discussed previously.
Second, $\bar{\eta}^{abc}_{L,1/2}$ are symmetric with respect to indices $b$ and $c$
since the injection DTN is excited by LPL \cite{HWang2020}.
Third, we find that $\bar{\eta}^{abc}_{L,1/2}$ are $\mathcal{T}$-even tensors, since
$\mathcal{T}v^a_{n}=-v^a_{n}$ and $\mathcal{T}g^{bc}_{nl}=g^{bc}_{ln}$,
where $\mathcal{T}$ denotes the time-reversal symmetry.
Finally, for the two-band Dirac Hamiltonian Eq.~(\ref{Hamiltonian}),
we find that Eqs.~(\ref{etaL1}-\ref{etaL2}) can be similarly rewritten as
\begin{align}
\bar{\eta}^{abc}_{L,1} = \mathcal{C}(\omega) \mathcal{N}^{abc}_1,
\quad
\bar{\eta}^{abc}_{L,2} = \mathcal{D}(\omega) \mathcal{N}^{abc}_2,
\label{DN}
\end{align}
where $\mathcal{C}(\omega)$ and $\mathcal{D}(\omega)$
have been defined in Eq.~(\ref{simp}) and
\begin{align}
\mathcal{N}^{abc}_1 & \equiv
\int_k \Xi^{abc}_{-+} \delta(\omega-2d),
\label{etaL11}
\\
\mathcal{N}^{abc}_2 & \equiv
\int_k \Lambda_{-+}^{abc} \delta(\omega-2d).
\label{etaL22}
\end{align}
Note that the quantum metric $g^{ab}_{-+}$ \cite{quantummetric},
appearing in both $\Xi^{abc}_{-+}$ and $\Lambda^{abc}_{-+}$,
plays a critical role in ensuring a nonvanishing \textit{anomalous} DTN,
as will be illustrated below. In addition, 
$\Xi^{abc}_{nm}E^bE^c=N^{(1)}_{nm}$,
$\Lambda^{abc}_{nm}E^bE^c=N^{(2)}_{nm}$,
$\mathcal{N}_1^{abc}E^bE^c=\mathcal{N}^{(1)}$,
and
$\mathcal{N}_2^{abc}E^bE^c=\mathcal{N}^{(2)}$.

\begin{figure}[t!]
\includegraphics[width=8.0cm]{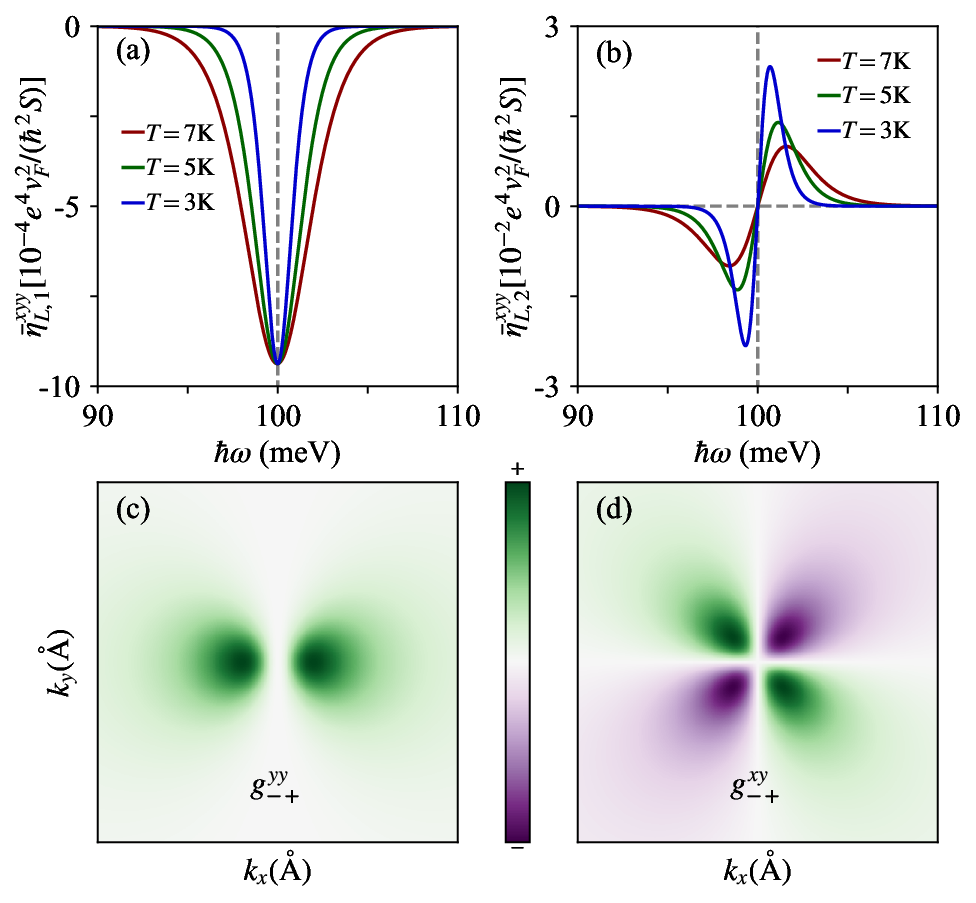}
\caption{
The temperature and frequency dependencies of
(a) $\bar{\eta}_{L,1}^{xyy}$ and (b) $\bar{\eta}_{L,2}^{xyy}$,
as calculated in Eqs.~(\ref{result2D1}-\ref{result2D2}), respectively.
Note that the vertical axis between (a) and (b) differs by a factor of $10^{2}$.
Here we have also set $\mu=50 \mathrm{meV}$.
The $\vect{k}$-resolved quantum metric (c) $g_{-+}^{yy}$ and (d) $g_{-+}^{xy}$.
}
\label{FIG2}
\end{figure}

\bigskip
\noindent{\textit{\textcolor{blue}{\textit{Anomalous} DTN in 2D.}}} ---
We first explore the \textit{anomalous} DTN in 2D monolayer graphene,
where $\vect{d}=(k_x, k_y,0)$ in Eq.~(\ref{Hamiltonian}) to capture
its low-energy physics near $\vect{K}$ or $\vect{K}'$ point \cite{Xiao2007graphene}.
For this model, $\omega_{+-}=2k$ with $k=(k_x^2+k_y^2)^{1/2}$,
$v_{\pm}^a=\pm k_a/k$, and $g^{bc}_{\mp \pm}=(k^2\delta_{bc}-k_bk_c)/(2k^4)$.
Using these expressions, along with Eqs.~(\ref{etaL11}-\ref{etaL22}) the first and second \textit{anomalous} DTNs can be calculated as
\begin{align}
\bar{\eta}^{abb}_{L,1}
&=
3\bar{\eta}^{aaa}_{L,1}
=
-\frac{e^4}{\hbar^2}
\frac{v_F^2}{S}
\frac{3\mathcal{C}(\omega)}{8\omega},
\quad a, b \in \{x, y\},
\label{result2D1}
\\
\bar{\eta}^{abb}_{L,2}
&=
-
\bar{\eta}^{aaa}_{L,2}
=
\frac{e^4}{\hbar^2}
\frac{v_F^2}{S}
\frac{\mathcal{D}(\omega)}{16\omega},
\quad a, b \in \{x, y\},
\label{result2D2}
\end{align}
where $e$, $\hbar$, and $v_F$ (the Fermi velocity) are restored by dimension analysis
and $S$ is the area of the system under investigation.

As expected, the resonant behavior of $\bar{\eta}^{abb/aaa}_{L,1}$ and $\bar{\eta}^{abb/aaa}_{L,2}$,
is fully determined by the dimensionless weights $\mathcal{C}(\omega)$ and $\mathcal{D}(\omega)$, respectively,
as illustrated in Figs.~\ref{FIG2}a-\ref{FIG2}b for the $xyy$ component,
in which the corresponding quantum metric are also shown [see Figs.~\ref{FIG2}c-\ref{FIG2}d].
While both $\bar{\eta}^{xyy}_{L,1}$ and $\bar{\eta}^{xyy}_{L,2}$ survive as $T \rightarrow 0$,
we observe that the peak value of $\bar{\eta}^{xyy}_{L,2}$
is two orders of magnitude larger than that of $\bar{\eta}^{xyy}_{L,1}$ at $T < 10 \mathrm{K}$.
This difference grows as $T \rightarrow 0$,
since the resonant peak of $\bar{\eta}^{xyy}_{L,2}$ scales as $1/T$
whereas the peak value of $\bar{\eta}^{xyy}_{L,1}$ remains intact at $\omega=2|\mu|$ when $T\rightarrow 0$.
Thus, Eq.~(\ref{etaL11}) can be safely neglected
at low temperatures (such as $T<10\mathrm{K}$),
and Eq.~(\ref{etaL22}) becomes the dominant contribution to the \textit{anomalous} DTN.

In addition to monolayer graphene, similar physics can be observed in
the surface state of 3D topological insulators,
which are effectively described by $\vect{d}=(-k_y, k_x, 0)$ \cite{Fu2009}.
Since this model has the same dispersion and quantum metric as the monolayer graphene,
we arrive at the same physics discussed above.
Beyond the linear dispersion, Eq.~(\ref{Hamiltonian}) can also
include quadratic terms on $\vect{k}$. For example,
it is straightforward to show that the Hamiltonian $H = k^2 + \lambda \vect{d} \cdot \vect{\sigma}$,
where $\vect{d}$ depends linearly on $\vect{k}$, can also give rise to \textit{anomalous} DTN.
This Hamiltonian in general describes three typical 2D systems with spin-orbit coupling (SOC) \cite{TaoLL}:
(1) Rashba SOC; (2) Dresselhaus SOC; and (3) Weyl SOC.

\bigskip
\noindent{\textcolor{blue}{\textit{\textit{Anomalous} DTN in 3D.}}} ---
In 3D, we consider the low-energy effective model of Weyl semimetals,
described by $\vect{d}=(k_x, k_y, k_z)$ near the Weyl node \cite{Weyl1}.
For this model, $\omega_{+-}=2h$ where $h=(k_x^2+k_y^2+k_z^2)^{1/2}$,
$v_{\pm}^a=\pm k_a/k$, and $g^{bc}_{\mp\pm}=(k^2\delta_{bc}-k_bk_c)/(2h^4)$. Using these expressions along with
Eqs.~(\ref{etaL11}-\ref{etaL22}) we find
\begin{align}
\bar{\eta}^{abb}_{L,1}
&=
2\bar{\eta}^{aaa}_{L,1}
=
-
\frac{e^4}{\hbar^2}
\frac{v_F}{V}
\frac{2\mathcal{C}(\omega)}{15\pi},
a,b,c \in \{x, y, z\},
\label{result3D1}
\\
\bar{\eta}^{aaa}_{L,2}
&=
-
2\bar{\eta}^{abb}_{L,2}
=
-\frac{e^4}{\hbar^2}
\frac{v_F}{V}
\frac{\mathcal{D}(\omega)}{30\pi},
a,b,c \in \{x, y, z\},
\label{result3D2}
\end{align}
which also shows the resonant behavior determined by the dimensionless weights
$\mathcal{C}(\omega)$ and $\mathcal{D}(\omega)$,
respectively. As a result, both Eq.~(\ref{result3D1}) and Eq.~(\ref{result3D2})
remain valid even at zero temperature, with Eq.~(\ref{result3D2})
becoming the dominant contribution at low temperatures
(e.g. $T < 10 \mathrm{K}$).

\bigskip
\noindent{\textit{\textcolor{blue}{\textit{Anomalous} DTN versus DSN.}}} ---
In mesoscopic conductors, the shot noise (SN) dominates over the
thermal noise (TN) at low temperatures. This naturally raises a question:
Can the \textit{anomalous} DTN be overshadowed by the DSN in bulk crystals even under light illumination?
In metallic crystals with time-reversal symmetry,
the DSN also consists of two distinct contributions,
which, within the two-band limit, are given by \cite{sup,note10}:
\begin{align}
\eta_{L,1}^{abc} &=
\int_k f_{nm}^2 \Xi^{abc}_{nm} \delta(\omega-\omega_{mn}),
\label{DSNSea}
\\
\eta^{abc}_{L,2}
&=
\int_k
\omega \dfrac{\partial f_{nm}^2}{\partial \epsilon_n}
\Lambda^{abc}_{nm}
\delta(\omega-\omega_{mn}),
\label{DSNsurf}
\end{align}
where the first term $\eta^{abc}_{L,1}$ with $f_{nm} \equiv f_n-f_m$ stands for
the Fermi-sea contribution while the second term $\eta^{abc}_{L,2}$ involving $\partial f_{nm}^2/\partial \epsilon_n$
accounts for the Fermi-surface contribution.

\begin{figure}[t!]
\includegraphics[width=8.0cm]{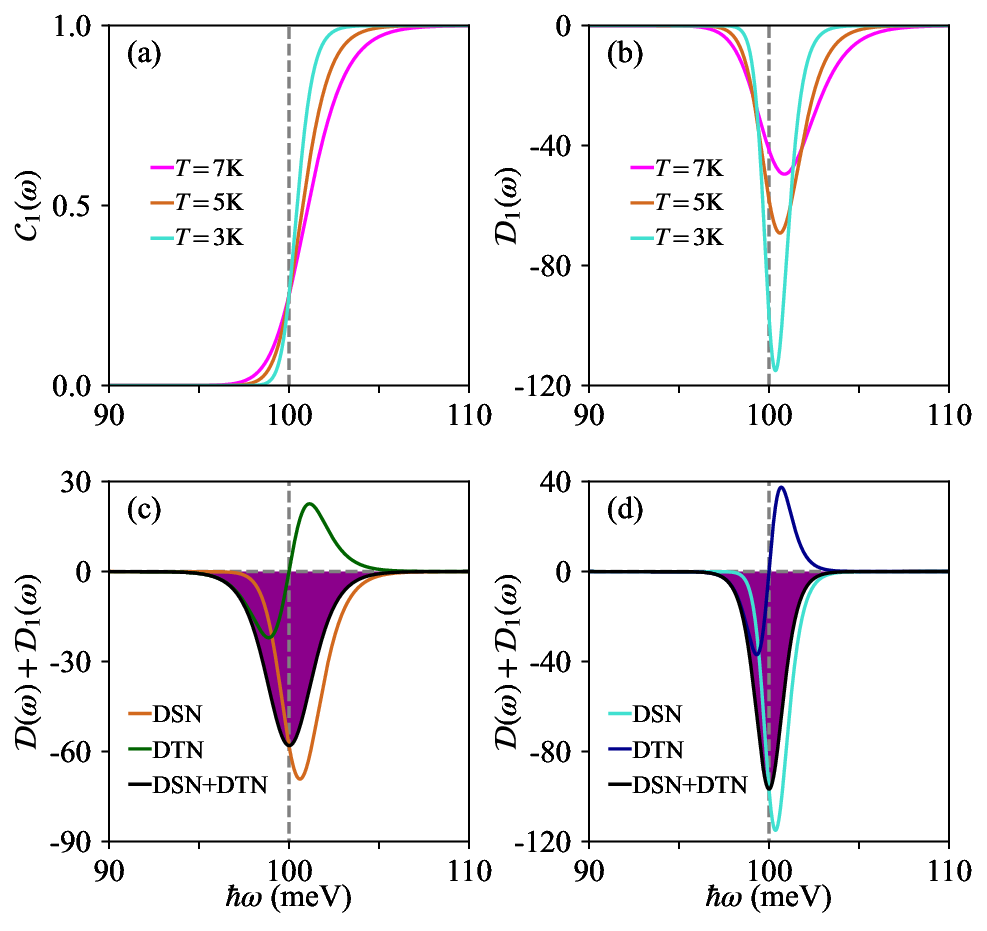}
\caption{
The temperature and frequency dependencies of (a) $\mathcal{C}_1(\omega)$
and (b) $\mathcal{D}_1(\omega)$, as defined in Eqs.~(\ref{C1}-\ref{D1}).
The total injection noise spectrum contributed by the \textit{anomalous} DTN
and the DSN at (c) $T=5\mathrm{K}$ and (d) $T=3\mathrm{K}$.}
\label{FIG3}
\end{figure}

Similarly, for Eq.~(\ref{Hamiltonian}), by using the optical selection rule $\delta(\omega-\omega_{mn})$
to factorize out the electron occupation information, we similarly obtain
\begin{align}
\eta^{abc}_{L,1}=\mathcal{C}_1(\omega) \mathcal{N}^{abc}_1,
\quad
\eta^{abc}_{L,2}=\mathcal{D}_1(\omega) \mathcal{N}^{abc}_2,
\label{DN1}
\end{align}
where $\mathcal{N}^{abc}_{1/2}$ have been defined in Eqs.~(\ref{etaL11}-\ref{etaL22}),
$\mathcal{C}_1(\omega) \equiv f_{-+}^2$
and $\mathcal{D}_1(\omega) \equiv \omega \partial f_{-+}^2/\partial \epsilon_{-}$
are also dimensionless weights. Explicitly, they are given by \cite{sup}
\begin{align}
\mathcal{C}_1 (\omega) & = \frac{\sinh^2(l_1/2)}{\left[\cosh(l_2)+\cosh(l_1/2) \right]^2},
\label{C1} \\
\mathcal{D}_1 (\omega) & = -l_1 \frac{2\sinh(l_1/2)+\sinh(l_1)\cosh(l_2)}{\left[ \cosh(l_2) + \cosh(l_1/2) \right]^3},
\label{D1}
\end{align}
where $l_1=\hbar\omega/(k_BT)$ and $l_2=\mu/(k_BT)$ with $\hbar$ and $k_B$ being restored by dimension analysis.
By comparing $\mathcal{C}_1(\omega)$ with $\mathcal{D}_1(\omega)$,
as shown in Figs.~\ref{FIG3}a-\ref{FIG3}b for various temperatures,
we find that the Fermi-surface contribution of the DSN
is one to two orders of magnitude larger than its Fermi-sea contribution at low temperatures
(e.g., $T < 10 \mathrm{K}$)
since $\mathcal{N}^{abc}_{1}$ and $\mathcal{N}^{abc}_2$ are of the same order.
Therefore, the Fermi-surface contribution becomes
the dominant source of the DSN at low temperatures (e.g., $T < 10 \mathrm{K}$).
Interestingly,
we note that the Fermi-surface contribution of DSN also features a $1/T$ divergence,
similar to the second type of \textit{anomalous} DTN.
Note that the $1/T$-DSN can also not be expected in mesoscopic conductors
or in bulk crystals under a static electric field.

Further, by comparing Eq.~(\ref{DN1}) with Eq.~(\ref{DN}),
we find that the difference between the \textit{anomalous} DTN and the DSN
at low temperatures is completely decided by their dimensionless weights [$\mathcal{D}(\omega)$ versus $\mathcal{D}_1(\omega)$].
Essentially, this is because both DSN and DTN are derived from Eq.~(\ref{Sab})
so that $\eta^{abc}_{L,1}$ and $\eta^{abc}_{L,2}$ can be obtained
by replacing $\bar{f}_{nm}$ with $f_{nm}^2$ in $\bar{\eta}^{abc}_{L,1}$ and $\bar{\eta}^{abc}_{L,2}$,
respectively, as adopted in Supplemental Material \cite{sup}.
Notably, we find that $\mathcal{D}(\omega)$ and $\mathcal{D}_1(\omega)$ are of the same order at a given temperature
and therefore the \textit{anomalous} DTN can not be overshadowed by the DSN,
as is typically observed in mesoscopic conductors.

Finally, we wish to mention that the resonant behaviors of $\mathcal{D}(\omega)$ and $\mathcal{D}_1(\omega)$ are markedly different.
Specifically, $\mathcal{D}_1(\omega)$ exhibits a single resonant peak at the right side of $\omega=2|\mu|$,
whereas $\mathcal{D}(\omega)$ shows resonant peaks on both side of $\omega=2|\mu|$,
particularly in an antisymmetric pattern.
As a result, the total injection noise spectrum at low temperatures contributed by both the \textit{anomalous} DTN
and the $1/T$-DSN, which is defined by
$\eta_{tot}=\left[\mathcal{D}(\omega)+\mathcal{D}_1(\omega)\right]\mathcal{N}_2^{abc}$
universally shows a resonant peak at $\omega_p=2|\mu|$,
as illustrated in Figs.~\ref{FIG3}c-d for $T=5\mathrm{K}$ and $T=3\mathrm{K}$, respectively.
Once the total noise spectrum is experimentally measured,
the DTN and DSN contributions can be isolated as
\begin{align}
\bar{\eta}_{L,2} =\dfrac{\mathcal{D}}{\mathcal{D} + \mathcal{D}_1}\eta_{tot},
\quad
\eta_{L,2} =\dfrac{\mathcal{D}_1}{\mathcal{D} + \mathcal{D}_1}\eta_{tot},
\end{align}
respectively, where $\mathcal{D}$ and $\mathcal{D}_1$ are dimensionless universal functions
decided by the unique electron-photon interaction.

\bigskip
\noindent{\textcolor{blue}{\textit{Summary}.}} ---
We developed the quantum theory of DTN in bulk crystals under light illumination, demonstrating that
in time-reversal-invariant systems, the DTN is closely connected to the quantum metric.
Remarkably, the DTN can exhibit a $1/T$ thermal noise \textit{anomaly},
arising from the interplay between the optical selection rule and the inherent Fermi-surface property of DTN.
Our findings are illustrated using massless Dirac or Weyl Hamiltonians.
Instead of being overshadowed by DSN, the \textit{anomalous} DTN is comparable with the DSN and
can lead to a universally resonant behavior of the total noise spectrum at low temperatures.
We wish to remark that the shot noise generated by nonequilibrium electrons
will produce fluctuating electromagnetic evanescent fields at the surface of the material. These fields
can be detected non-invasively using the scanning noise microscope \cite{exp1, exp2},
without the need for metallic electrodes typically used in mesoscopic measurements \cite{exp3, exp4}.
These experimental progresses provide a potential pathway to verify our proposals in Dirac materials.

\section*{Acknowledgements}
This work was financially supported by the Natural Science Foundation of China
(Grants No.12034014, No.12474047, and No.12174231.)


\begin{thebibliography}{00}
\bibitem{transitionrule}
The optical selection rule dictates that the incident light frequency $\omega$ must match
the energy difference $\hbar\omega_{nm}=\epsilon_n-\epsilon_m$ between
the $n$th band and the $m$th one to allow the optical transition.
\bibitem{Todd}
As a comparison, we wish to remark that the shift and injection DTNs,
excited by LPL and CPL, respectively,
are related to the local Berry curvature defined by $\Omega^{bc}_{nm}=i(r^b_{nm}r^c_{mn}-r^c_{nm}r^b_{mn})$
and the group velocity difference $\Delta^a_{nm}$ defined in Eqs.~(\ref{etaL1}-\ref{etaL2}),
in a way of $\Delta^a_{nm}(\Omega^{ab}_{nm}\Delta^c_{mn}+\Omega^{ac}_{nm}\Delta^b_{mn})$ and $(\Delta^a_{nm})^2\Omega_{nm}^{bc}$.
\bibitem{note10}
We use ${\bar \eta}$ and $\eta$ to denote, respectively, the injection DTN and DSN.
\end{thebibliography}
\end{document}